\documentclass{ws-p10x7}

\usepackage{amssymb}
\usepackage{graphicx}
\usepackage{axodraw}

\newcommand{\be}{\begin{equation}}
\newcommand{\ee}{\end{equation}}
\newcommand{\ba}{\begin{eqnarray}}
\newcommand{\ea}{\end{eqnarray}}

\newcommand{\rcite}[1]{$^{\ref{#1}}$}
\newcommand{\rbibitem}[1]{\bibitem{#1}\label{#1}}

\begin{document}

\title{\hfill\hfill
Chiral Lagrangians$^{*}$\hfill{\normalsize\rm LU TP 01-26\\
\hfill \lowercase{hep-ph}/0108111\\
\hfill A\lowercase{ugust 2001}\vskip-1.5cm}}

\author{Johan Bijnens}

\address{Department of Theoretical Physics, Lund University,\\
Solvegatan 14A, S 22362 Lund, Sweden\\
E-mail: bijnens@thep.lu.se\vskip-0.3cm}

\twocolumn[\maketitle\abstract{
An overview of the field of Chiral Lagrangians is given. This includes
Chiral Perturbation Theory and resummations to extend it to higher
energies, applications to the muon anomalous magnetic moment,
$\epsilon^\prime/\epsilon$ and others.
}]

\section{Introduction}

\footnotetext{$^*$Invited talk at the 
XX International Symposium on Lepton and Photon Interactions at High Energies
23rd-28th July 2001, Rome Italy.}
Chiral Symmetry is important in a lot of situations. In this talk I will
restrict myself to consequences of Chiral Symmetry for the strong
interaction. The subject is very broad as can be judged from
the lectures and review articles\cite{CHPTlectures}. In its modern form it
was founded by Weinberg\cite{Weinbergphysica} and 
Gasser and Leutwyler.\cite{GL1,GL2} I will discuss a few of the basics
in sects. \ref{chiralsym}, \ref{uses}, \ref{CHPT}. The applications
to $\pi\pi$ scattering (sect. \ref{sectpipi}), some two-flavour
results (sect. \ref{sectFV}) and resummations, and the question
of hadronic contributions to the muon magnetic moment (sect. \ref{sectamu}),
follow. Some of the theoretical developments in the structure and
understanding in particular of the many free parameters
follow in sect. \ref{structure}. Applications to three flavours, 
sect. \ref{3fl}, quark mass ratios, sect. \ref{quarkmass},
$V_{us}$, sect. \ref{sectvus}, anomalies and eta decays,
sects. \ref{anomalies}, \ref{moreeta}, semileptonic, sect. \ref{semi},
and nonleptonic, sect. \ref{nonl} weak decays 
and $\epsilon^\prime/\epsilon$ form the main remaining part.
I then conclude by a solar and cosmological as well as a high density
application together with some references to neglected areas.

More than
300 papers cited one of the three seminal papers in the last two
years, obviously necessitating many omissions.

\section{Chiral Symmetry}
\label{chiralsym}

QCD with 3 light quarks of equal mass has an obvious symmetry under
continuous interchange of the quark flavours. This is the well
known $SU(3)_V$. However, for $m_q=0$, 
\ba
{\cal L}_{QCD} &=&  
\sum_{q=u,d,s}
\left[i \bar q_L D\hskip-1.5ex/\, q_L +i \bar q_R D\hskip-1.5ex/\, q_R\right.
\nonumber\\&&\left.
- m_q\left(\bar q_R q_L + \bar q_L q_R \right)
\right]
\ea
has as symmetry the full chiral $SU(3)_L\times SU(3)_R$
since the left and
right handed quarks decouple.
Massive particles can always be changed from left to right handed
by going to a Lorentzframe that moves faster than the particle, this changes
the momentum direction but not the spin direction and hence flips
the helicity.\footnote{For Majorana masses this Lorentz transformation also
changes particle into antiparticle.} For massless particles this argument fails
and the left and right helicities can thus be rotated separately.

Chiral Symmetry is broken by the vacuum of QCD, otherwise we would
see a parity partner of the proton at a similar mass. Instead we believe
that this symmetry is spontaneously broken by a quark-antiquark condensate
or vacuum-expectation-value (VEV)
\be
\label{vev}
\langle \bar q q\rangle = \langle \bar q_L q_R+\bar q_R q_L\rangle
\ne 0\,.
\ee
This condensate breaks $SU(3)_L\times SU(3)_R$ down to the diagonal
subgroup $SU(3)_V$. This breaks 8 continuous symmetries and we must
thus have 8 massless particles, Goldstone Bosons, whose interactions
vanish at zero momentum. 

The VEV that spontaneously breaks the Chiral Symmetry can also
be a different one than in Eq. (\ref{vev}). This option
is often known as Generalized Chiral Perturbation Theory.\cite{GCHPT}

\section{Uses of Chiral Symmetry}
\label{uses}

Chiral Symmetry can be used in high energy and nuclear physics in
a variety of ways:
\begin{itemize}
\item Constructing {\em chirally invariant phenomenological Lagrangians}
 to be used only at tree level.
\item {\em Current Algebra} which directly uses the Ward Identities of
$SU(3)_L\times SU(3)_R$ and the Goldstone Boson nature of the pion to restrict
amplitudes. Often these calculations assume smoothness assumptions on
the amplitudes. This method is very powerful but becomes unwieldy when going
beyond the leading terms.
\item {\em Chiral Perturbation Theory} (CHPT) which is the modern
implementation of current algebra using the full power of 
{\em effective field theory} (EFT) methods. In recent years CHPT methods
have been developed for most areas where current algebra is applicable
in particular for mesons with two and three flavours, single baryons,
two or three baryons, and also in including nonleptonic weak and
electromagnetic interactions.
\item Using {\em dispersion relations} with CHPT constraints as a method
to include higher orders and/or extend the range of validity
of the CHPT results.
\item The use of all the above in estimating weak nonleptonic decays
and in particular $\epsilon^\prime/\epsilon$. 
\end{itemize}

\section{Chiral Perturbation Theory}
\label{CHPT}

As degrees of freedom we use the eight Goldstone Bosons of spontaneous
chiral symmetry breaking, identified with the $\pi,K,\eta$ octet,
and we expand in momenta using the fact that
the interaction vanishes at zero momentum.
The precise procedure can be found in many lectures\cite{CHPTlectures}
but is referred to as powercounting in generic momenta ($p$),
external currents
and quark masses. The usual ordering is $m_q\sim p^2$ since
$m_\pi^2\sim p^2\sim m_q\langle\bar q q\rangle$ and we count external photon
$Z^0$ and $W^\pm_\mu$ fields as order $p$ since they occur together
with a momentum in the covariant derivative
\be
D_\mu = \partial_\mu -ieA_\mu\,.
\ee
An example for the powercounting in $\pi\pi$-scattering is shown below:
\\[0.2mm]
\setlength{\unitlength}{0.5pt}
\SetScale{0.5}
\begin{picture}(100,100)
\SetWidth{1.5}
\Line(0,100)(100,0)
\Line(0,0)(100,100)
\Vertex(50,50){5}
\end{picture}
\raisebox{25pt}{ Meson Vertex}\hfill
\raisebox{25pt}{$p^2$}\\[0.5cm]
\begin{picture}(100,30)
\SetWidth{1.5}
\Line(0,15)(100,15)
\end{picture}
\raisebox{5pt}{ meson propagator}\hfill\raisebox{10pt}{$1/p^2$}\\[0.5cm]
$\int d^4p$\quad loop integral\hfill$p^4$\\[0.5cm]
\begin{picture}(100,100)
\SetWidth{1.5}
\Line(0,100)(20,50)
\Line(0,0)(20,50)
\Vertex(20,50){5}
\CArc(50,50)(30,0,180)
\CArc(50,50)(30,180,360)
\Vertex(80,50){5}
\Line(80,50)(100,100)
\Line(80,50)(100,0)
\end{picture}
\hfill\raisebox{25pt}
{$(p^2)^2\,(1/p^2)^2\,p^4 = p^4$}\\[0.5cm]
\begin{picture}(100,100)
\SetWidth{1.5}
\Line(0,0)(50,40)
\Line(0,50)(50,40)
\CArc(50,70)(30,0,180)
\CArc(50,70)(30,180,360)
\Vertex(50,40){5}
\Line(50,40)(100,50)
\Line(50,40)(100,0)
\end{picture}
\hfill\raisebox{25pt}
{$(p^2)\,(1/p^2)\,p^4 = p^4$}
\\
The lowest order diagram is just the tree level vertex at $p^2$
and as can be seen the two one-loop diagrams are both $p^4$. The existence
of this loop expansion was shown in a very nice paper
by Weinberg.\cite{Weinbergphysica} This paper can really be considered the
birth of modern CHPT.

\section{$\pi$-$\pi$ scattering}
\label{sectpipi}

The amplitude for $\pi$-$\pi$ scattering can, in isospin notation, be written
as
$$
A(\pi^i\pi^j\to \pi^k\pi^l) = i\delta^{ij}\delta^{kl}A(s,t,u)
+\mbox{ cyclic}\,.
$$
The fact that $\pi$-$\pi$ scattering is weak near threshold
is one of the major qualitative predictions of spontaneous chiral
symmetry breaking. The order $p^2$ contribution was worked out using
current algebra methods by Weinberg in the sixties:\cite{Weinbergpipi}
\be
 A(s,t,u) = 
(s-m_\pi^2)/F_\pi^2\,.
\ee
The $p^4$, including loop-diagrams and the new free parameters, was done by
Gasser and Leutwyler\cite{GLpipi} and the full two-loop expression
was performed recently\cite{pipitwoloop} after partial calculations
in\cite{pipidispersive}. Remarkably the full amplitude can be written
in terms of logarithms and other elementary functions.

The new results in the last year are that the Roy equation analysis
was updated using the larger computer power now available\cite{pipiRoy}
and the new data from the BNL-E865 experiment\cite{BNLE865}.
Both of these were then combined with the two-loop CHPT
results in \cite{pipifinal} with the conclusions that standard CHPT
works find and that, at least in the two-flavour sector, the more generalized
scenario (GCHPT) is not needed.

\begin{figure}
\includegraphics[width=\columnwidth]{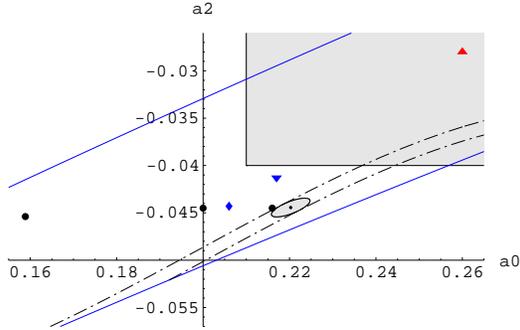}
\caption{\label{fig1pipi} The conclusions of the new data on $\pi\pi$
scattering analyzed using CHPT and Roy equations. The
triangle, $\blacktriangle$, and the shaded region are the old data.
The band shows the Roy constraints and $\blacklozenge,\blacktriangledown$
are two predictions of pure CHPT indicating the uncertainty
in the parameters.
The dash-dotted band is the Roy equation analysis including (G)CHPT
constraints
and the ellipse shows the result of the new data.
The bullets from left to right show the convergence of standard CHPT
at orders $p^{2,4,6}$. From\rcite{pipifinal}.}
\end{figure}

\begin{figure}
\includegraphics[width=\columnwidth]{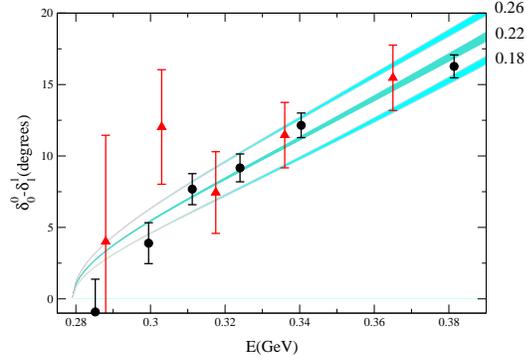}
\caption{\label{fig2pipi}
$\bullet$ are the new  BNL E865 data and $\blacktriangle$
the older Rosselet\rcite{Rosselet} ones. The three bands are
the predictions for three different values of the scattering length $a_0^0$.
Figure from \rcite{pipifinal}.}
\end{figure}

In Fig.~\ref{fig1pipi} we show their conclusions
and in Fig.~\ref{fig2pipi} the agreement with the old and the new data.
In the future we expect a further improvement from measurements in $K_{e4}$
at KLOE and from pionium atoms at DIRAC.\cite{DIRAC}
The theoretical calculations needed for the latter data were
recently completed.\cite{Gasserpionium}

\section{Other 2 flavour CHPT}
\label{sectFV}

The first two-loop calculation in CHPT was the two-flavour
process $\gamma\gamma\to\pi^0\pi^0$ and its polarizabilities\cite{ggpp}
and the equivalent calculation for the charged pions.\cite{Buergi}
The latter also included the pion mass and decay-constants, 
see also.\cite{pipitwoloop}

Radiative pion decay, $\pi\to\ell\nu\gamma$, is also known.\cite{BTpilng}
The most recent calculation in this sector was the full CHPT calculation
of the pion scalar and vector form factors.\cite{piform}
There has since been quite some work trying to add dispersion theoretical
constraints to the pion vector form factor. Using inverse
amplitude methods and Omn\`es  equation inspired resummations
a very nice fit to the ALEPH\cite{ALEPH}
and CLEO-II\cite{CLEO} data for $\tau$-decay
was obtained. Similar work, with references to earlier work is \cite{Truong}.

\begin{figure}
\includegraphics[height=\columnwidth,angle=-90]{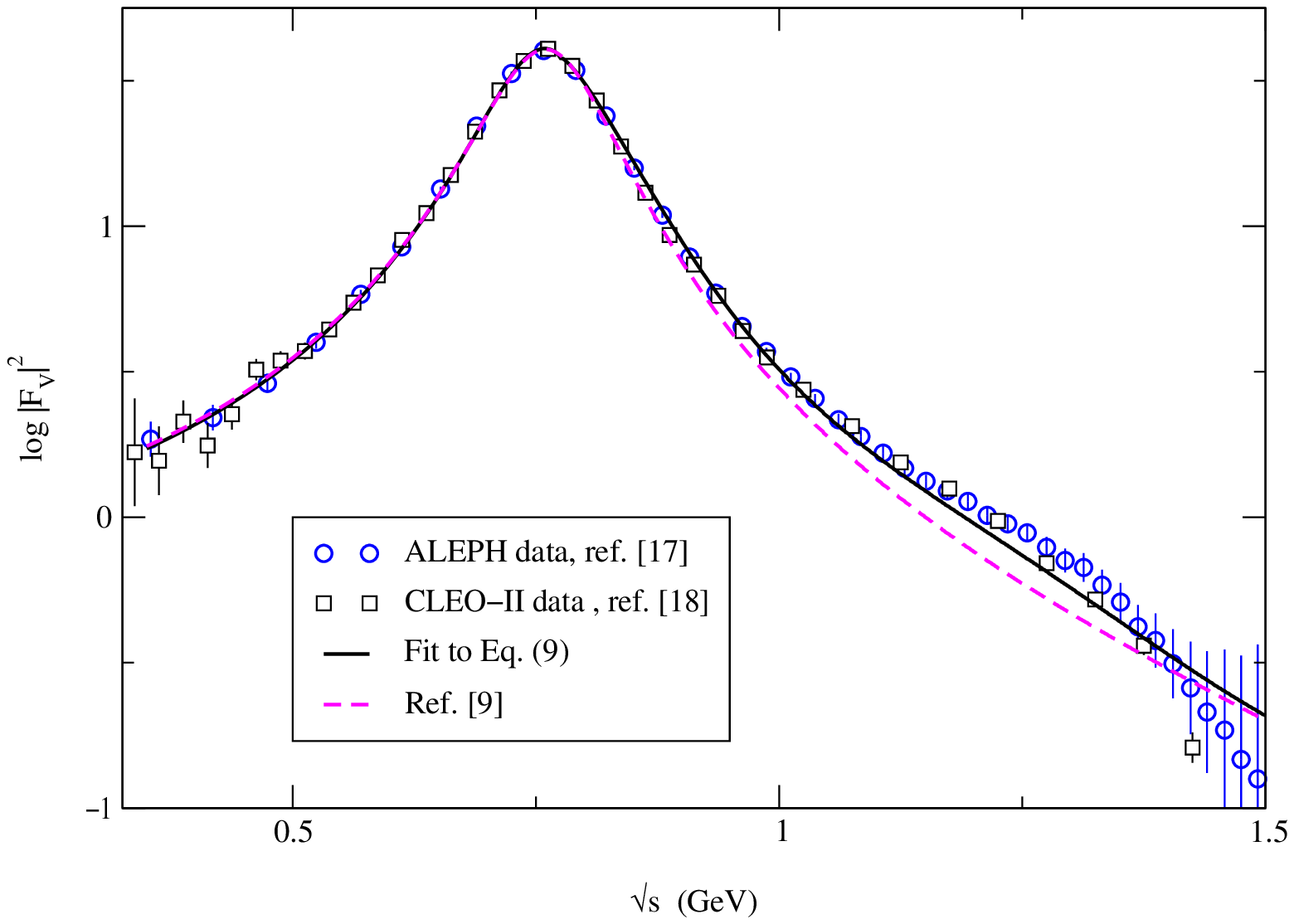}
\caption{\label{fig1FV} The resummed CHPT expression and the fit to
the $\tau\to\pi\pi\nu$ data. Figure from \rcite{PichPortoles}.}
\end{figure}

\begin{figure}
\includegraphics[height=\columnwidth,angle=-90]{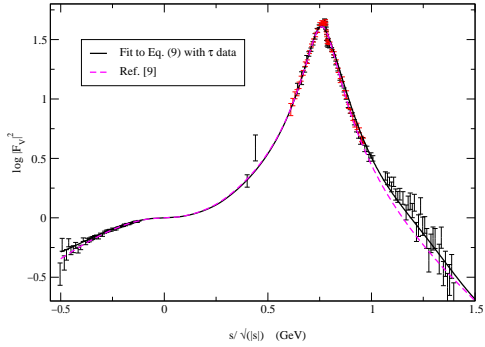}
\caption{\label{fig2FV} The resummed CHPT expression and the fit to
the $e^+e^-$ and spacelike data. Figure from \rcite{PichPortoles}.}
\end{figure}

In Fig. \ref{fig1FV} I show the quality of the fit to the $\tau$-data.
Good fits up to $\sqrt{s}\sim1.5~GeV$
are obtained also for the $e^+e^-\to\pi^+\pi^-$ data
and the spacelike $e\pi\to e\pi$ data\cite{NA7space}
as shown in Fig. \ref{fig2FV}.
The resummations in this sector work well, in Table \ref{table1FV}
I show the charge radius and the coefficient of the quartic $q^4$
term in the vector form factor from the dispersive fit\cite{PichPortoles}
and the pure CHPT fit.\cite{piform} Notice the quality of the agreement
and the similarity of the errors even though the dispersive fit is
dominated by the data around the rho peak and the CHPT fit by the
spacelike data.\footnote{The CHPT calculation has a large error
for $c_\pi$ due to the inclusion of rather incompatible
NA7 {\em timelike}-data.\cite{NA7time}}
\begin{table}
\begin{tabular}{|l|cc|}
\hline
           & dispersive & CHPT\\
\hline
$\langle r^2 \rangle^\pi_V$ (GeV $^{-2}$) & $11.04(30)$
                                          & $11.22(41)$\\
$c^\pi_V$ (GeV$^{-4}$)     & $3.79(4)$ 
                           & $3.85(60)$ \\
\hline
\end{tabular}
\caption{\label{table1FV} Comparison between the CHPT fits and the
dispersive improvement for the pion charge radius and the next term in
formfactor expansion.}
\end{table}

\section{$a_\mu$: muon magnetic moment}
\label{sectamu}

The measurement\cite{BNLgm2} is
\be
a_\mu = 116~592~023(151)~\times~10^{-11}
\ee
while a recent theory review quotes the standard model prediction\cite{CM}
\be
 a_\mu = 116~591~597(67)~\times~10^{-11}\,.
\ee
The difference is a few sigma depending on how errors are combined.
The size of the theory error has been disputed by many people, varying from
an ``I don't believe it'' to more reasonable partial studies,
an example of the latter is \cite{Melnikov}.

The methods of Chiral Lagrangians contribute to this theory prediction
in various ways
\begin{enumerate}
\item The very low energy vacuum polarization contribution.
\item The effects of isospin breaking, in particular the issue
of $\tau\to\pi\pi\nu$ data versus  $e^+e^-\to\pi\pi$ data.
\item The calculation of the hadronic contribution to
 light-by-light scattering.
\item The~ EFT~calculation~of~higher~order electroweak corrections.\cite{Peris}
\end{enumerate}
I now discuss the first three in more detail.

\subsection{$a_\mu$: vacuum-polarization}

\begin{figure}
\begin{center}
\includegraphics[width=0.7\columnwidth]{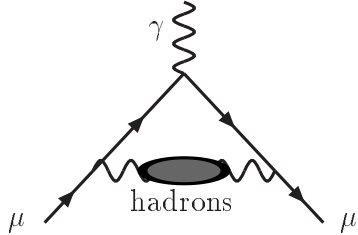}
\end{center}
\caption{\label{fig1amu} The hadronic vacuum polarization contribution to
the muon anomalous magnetic moment.}
\end{figure}

The hadronic vacuum polarization contribution\footnote{The QED corrections
are much larger but under good theoretical control.} is depicted in
Fig. \ref{fig1amu}.
The theory expression can be related to an integral over the
experimentally observable ratio of hadronic events to $\mu^+\mu^-$
pairs in $e^+e^-$ collisions:
\be
 a_\mu = \left(\frac{\alpha m_\mu}{3\pi}\right)^2
\int_{4m_\pi^2}^\infty dt\, R(t) \frac{\hat K(t)}{t^2}\,.
\ee
Here $\hat K(t)$ is a slowly varying function whose
expression can be found in many places.\cite{Jegerlehner}
The contributions at low energies are enhanced but due to the very
strong rho peak it is still dominated by that.

Pure CHPT methods can be used at $\sqrt{t}\le 0.5$~GeV.
Using the two-loop expression for the pion form factor
with all available low-energy data yields\cite{piform}
\be
\label{amuhadlow}
10^{11} a_\mu^{\pi\pi}(\sqrt{t}\le 0.5~GeV) = 563\pm30\,.
\ee
The error is mainly experimental, the best fit changes quite considerably
depending on whether the {\em timelike} NA7 data\cite{NA7time}
are included and the error on (\ref{amuhadlow}) reflects this.
 
The use of dispersion relations and resummations of the CHPT
result\cite{PichPortoles} allows to go higher in energy
leading to
\be
10^{11} a_\mu^{\pi\pi}(t\le 1.2~GeV) = 5113\pm60\,.
\ee
A recent more traditional evaluation also using the $\tau$-data
but employing similar theory constraints yields\cite{Yndurain}
\be
10^{11} a_\mu^{\pi\pi}(t\le 1.2~GeV) = 5004\pm52\,.
\ee
The other determinations do not quote the same energy range
for this contribution so cannot be compared directly, they are
discussed in the contribution by Miller. Note the difference between
the two estimates above.

\subsection{$\tau$ versus $e^+e^-$}

\begin{figure}
\includegraphics[width=\columnwidth]{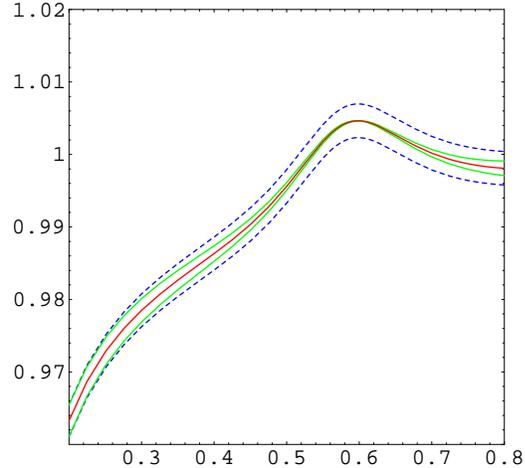}
\caption{The correction factor needed to go from $\tau$ to
$e^+e^-$ data as estimated in \rcite{CEN}. The $x$-axis is $t$ in units
of GeV$^2$, figure from \rcite{CEN}.}
\label{fig3amu} 
\end{figure}

\begin{figure}
\includegraphics[width=\columnwidth]{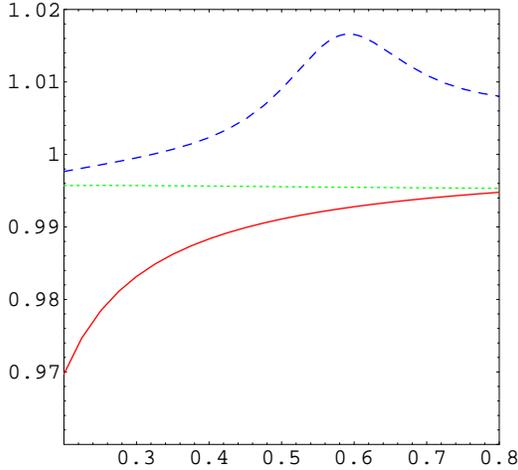}
\caption{The separate factors, the solid line is the kinematic
$\beta_{+-}^3/\beta_{00}^3$ and the dashed line the isospin corrections
to the form-factors. Figure from \rcite{CEN}.}
\label{fig4amu}
\end{figure}

One of the issues is whether isospin corrections in going from
 $\tau^+\to\pi^+\pi^0\bar\nu_\tau$ to
 $e^+e^-\to\pi^+\pi^-$ are large.
At the low-energy end, $\sqrt{t}\le 0.5$~GeV, CHPT can be used
to calculate these corrections.
Quark mass corrections are very small
since they are ${\cal O}\left((m_u-m_d)^2 \right)$.
The main effect comes from
photonic contributions.
A first evaluation of these
has been done recently.\cite{CEN} The CHPT calculations
are then extended to higher energies using the methods of \cite{PichPortoles}
discussed in Sect.~\ref{sectFV}. The result is a fairly small correction
factor shown in Fig. \ref{fig3amu}. The band is an indication of
the uncertainty.
The composition of the result is shown in Fig. \ref{fig4amu}.
More work on this correction is welcome.

\subsection{$a_\mu$: light-by-light}

\begin{figure}
\begin{center}
\includegraphics[width=0.7\columnwidth]{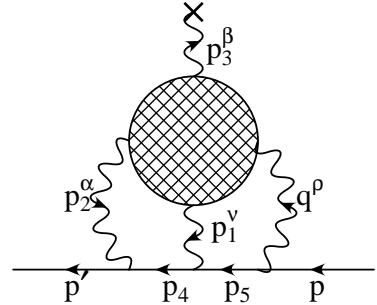}
\end{center}
\caption{The hadronic light-by-light contribution to the muon
anomalous magnetic moment.}
\label{fig5amu}
\end{figure}

The hadronic light-by-light contribution to the muon anomalous magnetic
moment is depicted in Fig. \ref{fig5amu}.
This contribution can be rewritten as an integral over 7 kinematic
variables. A simple relation to an integral of a measurable quantity
does not exist and given the large amount of kinematic variables,
the analytic structure of the underlying amplitude
is very complicated and makes a  dispersive  analysis nearly impossible.
We thus need a {\em pure} theory prediction.

We could start by a naive attempt and simply use a constituent quark loop.
This has been done in \cite{KNO85} with the result
\be
 a_\mu^{LL}(CQ) = 60(4)~10^{-11}\,.
\ee
How much can we trust these numbers ? Using the result for the
vacuum polarization
\be
\label{amuCQ}
a_\mu^q(CQ) = \alpha^2 Q_q^2 m_\mu^2/(15 \hat m_q^2 \pi^2)+\ldots
\ee
we obtain $\sim 2000~10^{-11}$ versus the result
of $\sim 7000~10^{-11}$ using the data and the dispersive
method.\footnote{We can of course fit the quark mass to obtain this result
but that is not a prediction.} We can thus obviously not trust the
result (\ref{amuCQ}) and need hadronic inputs.
This is not a CHPT calculation
as often stated, in pure CHPT we encounter divergent contributions
that necessitate a counterterm which is precisely the quantity we are trying
to predict.

The problem of doublecounting hadronic and quark-loop
contributions was alleviated very much by de Rafael\cite{Eduardo}
who noted that large $N_c$ counting and chiral powercounting could be used
a a guide to classify them.  He noted that the three main contributions are
\begin{enumerate}
\item  ${\cal O}(N_c)$ and $p^6$: $\pi^0,\eta,\eta^\prime$ exchange.
\item  ${\cal O}(N_c)$ and $p^8$: irreducible four-meson vertices
and exchanges of heavier resonances.
\item ${\cal O}(1)$ and $p^4$: $\pi^+,K^+$ loop\,.
\end{enumerate}

This method was then applied by two groups, 
Hayakawa and Kinoshita HK(S)\cite{Hayakawa} and 
Bijnens, Pallante and Prades (BPP)\cite{BPP}.
The results are shown in Table \ref{table1amu}.
\begin{table}
\begin{center}
\begin{tabular}{|l|cc|}
\hline
$10^{11}a_\mu$       &   BPP   &   HK(S) \\
\hline
1    &  $-$85(13)    &  $-$83(6) \\
2    &  12(6)     &   8(11)         \\
3   &  $-$19(13)    &  $-$4.5(8.1) \\
\hline
sum   & $-$92(32)      & $-$79(15) \\
\hline
\end{tabular}
\end{center}
\caption{The three classes of contributions to the hadronic light-by-light
for $a_\mu$ as obtained by the two groups.}
\label{table1amu}
\end{table}
For contribution 1, the two groups are in good agreement. Uncertainties
here are the choices of formfactors, in practice what both groups have chosen
amounts to double VMD, i.e., vector meson propagators are added in both
photon legs in the $\pi^0,\eta,\eta^\prime$ coupling to $\gamma\gamma$.
There are basically no data with both photons off-shell, we need
double tagged data at intermediate values of the photon off-shellness
and data on $\eta$ and $\eta^\prime$ decays to two lepton pairs,
i.e. on $\gamma^*\gamma^* \longleftrightarrow \pi^0,\eta,\eta^\prime$.
A preliminary study of the effects of various form factors on
$a_\mu$ and how they can be observed in the other experiments can be
found in \cite{Persson}, which yielded $a_\mu(1) = -88(9)~10^{-11}$.

For contribution 2, there are many small contributions with
alternating signs, including scalar and axial-vector exchange.

The third contribution is different in both approaches because
the choice of the underlying $\gamma^*\gamma^*\pi^+\pi^-$ vertex is
different, both choices satisfy chiral constraints. Both are possible and
this difference is at present inherent and provides a lower bound on the
error. We need information on
 $\gamma^*\gamma^*\to\pi^+\pi^-$ at intermediate to large off-shellness
for {\em both} photons to clarify this issue.

The matching of hadronic
and short-distance results was studied in an approximative
way in \cite{BPP} and found to be satisfactory.

Numerically the total contribution is dominated by the relatively well
understood pseudoscalar exchange. Given the complexity of the
underlying amplitude it is difficult to prove that no major
contribution has been missed but all additional effects
studied in \cite{Hayakawa,BPP} were relatively small.
The errors in \cite{BPP} were added linearly since all contributions
involve fairly similar assumptions and is in my opinion
reasonable. Unless a qualitatively different contribution from
the ones included in these two calculations is discovered I do not expect
the results to change significantly.

\section{Structure of CHPT}
\label{structure}
\subsection{History and overview}

The structure of effective Lagrangians for the pseudoscalar mesons was
originally worked out by Weinberg for two flavours and then generalized
to arbitrary symmetry groups.\cite{CCWZ}
Weinberg\cite{Weinbergphysica} introduced the full EFT formalism
which was then systematized and extended by Gasser and
Leutwyler.\cite{GL1,GL2} The number of parameters was two at tree level
and ten more at one-loop level.
A first attempt at classifying the next order was done in \cite{FS}
and later finished by \cite{BCE1}. The latter group also worked
out the full infinity structure at two-loop order.\cite{BCE2}

In the abnormal parity sector, including one power of
$\epsilon^{\mu\nu\alpha\beta}$, the lowest order is $p^4$ and is the
celebrated Wess-Zumino-Witten term with no
free parameters.\cite{WZ} The methods of going beyond lowest order
were worked out in \cite{Issleretal} and work is going on to determine
the precise number of parameters here.\cite{BGT}

The extension to the nonleptonic weak sector was done by Kambor
et al\cite{KMW} and to the quenched approximation
by Sharpe, Bernard and Golterman after early work by Morel.\cite{quenched}

Some indication of the amount of work done in this area is given in
Table \ref{table1chpt} where we indicated the
different Lagrangians people have considered and their number of parameters
in parentheses.
Recent lectures covering various aspects are \cite{CHPTlectures}.
\begin{table*}
\begin{center}
\begin{tabular}{|l|c|} 
\hline
\hspace{1cm} ${\cal L}_{\rm chiral\; order}$ 
~({ $\#$ of LECs})  &  loop order\\
\hline 
${\cal L}_{p^2}({ 2})$+${\cal L}_{p^4}^{\rm odd}({ 0})$+
${\cal L}_{G_Fp^2}^{\Delta S=1}({ 2})$ +${\cal L}_{e^2p^0}^{\rm em}({ 1})$
+${\cal L}_{G_8e^2p^0}^{\rm emweak}({ 1})$ 
+${\cal L}_{p}^{\pi N}({ 1})$+${\cal L}_{p^2}^{\pi N}({ 7})$ & $L=0$\\
+${\cal L}_{G_8p^0}^{MB,\Delta S=1}({ 2})$+
${\cal L}_{G_8p}^{MB,\Delta S=1}({ 8})$ 
+ ${\cal L}_{e^2p^0}^{\pi N,{\rm em}}({ 3})$ & \\
\hline
+${{\cal L}_{p^4}^{\rm even}({ 10})}$
+${{\cal L}_{p^6}^{\rm odd}({ 23})}$
+${{\cal L}_{G_8p^4}^{\Delta S=1}({ 22})}$
+${{\cal L}_{e^2p^2}^{\rm em}({ 14})}$ 
+${{\cal L}_{G_8e^2p^2}^{\rm emweak}({ 14})}$
+${{\cal L}_{e^2p}^{\rm leptons}({ 5})}$  & $L=1$\\
+${{\cal L}_{p^3}^{\pi N}({ 23})}$+
${{\cal L}_{p^4}^{\pi N}({ 114})}$
+${{\cal L}_{G_8p^2}^{MB,\Delta S=1}({ ?})}$ 
+${{\cal L}_{e^2p}^{\pi N,{\rm em}}({ 8})}$ 
 &  \\
\hline
+${{\cal L}_{p^6}^{\rm even}({ 90})}$  & $L=2$ \\
\hline
\end{tabular}
\end{center}
\caption{The standard model Lagrangian at low energies
or an overview of the various Chiral Lagrangians and their
number of parameters in parentheses, taken from \rcite{Ecker}.}
\label{table1chpt}
\end{table*}

\subsection{Parameter Estimates}

As can be seen from Table \ref{table1chpt} one of the major
problems in dealing with chiral Lagrangians is the number of free
parameters. The reason is that we only use
the chiral symmetry,  $SU(3)_L\times SU(3)_R$, and its
Goldstone character, all the remaining physics is parametrized.

First estimates of these parameters were done by
using resonance exchange.\cite{EGPR}
The conclusion is that the values of the $p^4$ parameters, $L_i^r$
are dominated by the vector and axial-vector degrees of freedom.
Whenever these dominate, predictions work well. In the scalar sector
qualitative agreement was obtained but the numerical agreement
was less accurate.
Later work concentrated on checking quark models in
particular the chiral quark model\cite{ERT}, the Nambu-Jona-Lasinio model
and extensions\cite{BBR}  and nonlocal quark models. Unfortunately,
in the latter case the number of free parameters becomes rather large
again.
More recent work has concentrated on including more
resonances\footnote{Examples of this are all the two-loop papers
cited elsewhere in this talk}  and a more systematic use of
short-distance constraints first started in\cite{EGLPR}.
Recent work is \cite{parameters}.

\section{CHPT in the three flavour sector}
\label{3fl}

Most basic two-loop calculations are done but a study of many smaller
processes remains to be done.
Finished ones include
the vector and axial-vector two-point functions, masses and decay
constants\cite{KG,ABT1}, the latter also including isospin
violation,\cite{ABT3} and the scalar two-point function\cite{Moussallam}.

The process $K\to\pi\pi\ell\nu$ is also known to this order,\cite{ABT2}
because it is needed to determine the parameters.
The results for the parameters that are known to two-loop order
is in Table \ref{table2chpt}. I have quoted here the results
from\cite{ABT3} using the new data\cite{BNLE865} rather than the
original fit\cite{ABT2} which only used the older data.\cite{Rosselet}
For pion and kaon vector formfactors partial results exist.\cite{PS,BT}
\begin{table*}
\begin{center}
\begin{tabular}{|c|cccccc|}
\hline
i   & 1 & 2 & 3 & 5 & 7 & 8 \\
\hline
$10^3 L_i^r(m_\rho)$ & $0.43(12)$ & $0.73(12)$ & $-2.35(37)$ &
$ 0.97(11)$ & $-0.31(14)$ & $0.60(18)$ \\
$p^4$ & 0.38 & 1.59 & $-2.91$ & 1.46 & -0.49 & 1.00\\
\hline
\end{tabular}
\end{center}
\caption{A fit to two-loop order of the $p^4$ CHPT parameters,
table from \rcite{ABT3}.}
\label{table2chpt}
\end{table*}
Some of the two-loop corrections are fairly sizable, especially in
the masses. The effect on mass ratios is smaller but claims have been
made that this indicates a GCHPT picture in the three flavour case.

\section{Quark Mass Ratios}
\label{quarkmass}

\begin{figure}
\includegraphics[height=\columnwidth,angle=-90]{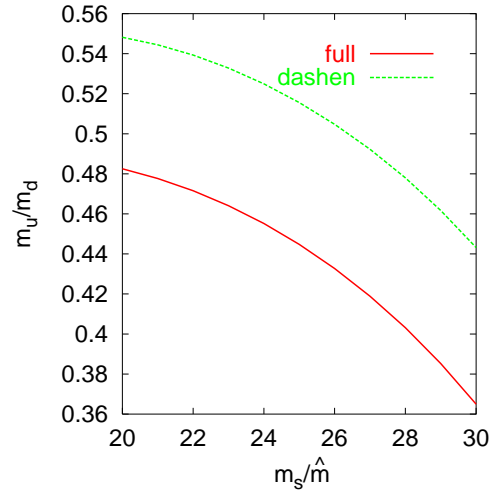}
\caption{The variation of $m_u/m_d$ with $2m_s/(m_u+m_d)$.
The top line is using the old estimate for the electromagnetic
Kaon mass difference. The bottom line uses the newer ones.
From \rcite{ABT3}.}
\label{fig1quark}
\end{figure}

One of the main results of the isospin breaking at two-loops so far is
a new determination of the quark mass ratio\cite{ABT3} $m_u/m_d$
\be
{m_u}/{m_d} = 0.46\pm0.09\,.
\ee
and the pion mass splitting from quark masses
$$
\left(m_{\pi^\pm}-m_{\pi^0}\right)_{\mbox{QCD}} = 0.32\pm0.20~
\mbox{MeV}\,.
$$
The uncertainty is larger than the usually quoted one
since at two-loop level the uncertainty due to choices of saturation scale
etc. were larger. We need at the present level $m_s/(m_u+m_d)$ as an input
parameter. The variation due to this is shown in Fig. \ref{fig1quark}.
The main reason of the change w.r.t. the old values is the much larger
estimate of the electromagnetic part of the Kaon mass
difference.\cite{dashen}

\begin{figure}
\includegraphics[height=\columnwidth,angle=-90]{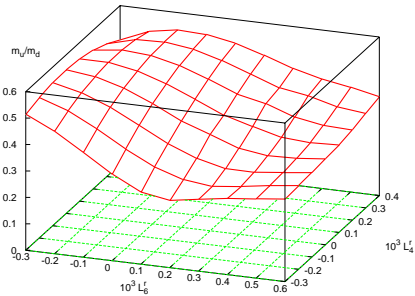}
\caption{The variation of $m_u/m_d$ with the large $N_c$ suppressed
input assumptions. Note that it never gets near zero,
from \rcite{ABT3}.}
\label{fig2quark}
\end{figure}

The uncertainty due to the Kaplan-Manohar ambiguity\cite{KM}
is fixed due to the limits on $L_4^r$ and $L_6^r$ obtained from
\cite{Moussallam}. Note that the values for $L_7^r$ obtained
are in perfect agreement with the arguments of \cite{mun0}
used to exclude the option $m_u=0$ there.
That $m_u/m_d$ never gets near zero is shown in Fig.~\ref{fig2quark}
where the range shown is significantly larger than allowed
by\cite{Moussallam}.

\section{Determination of $V_{us}$ (and $V_{ud}$)}
\label{sectvus}

The CKM matrix-element $V_{us}(V_{ud})$ can be determined from
the decays
$K\to\pi\ell\nu$ ($\pi^+\to\pi^0e^+\nu$)
and hyperon semileptonic decays (neutron or nuclear $\beta$ decays).
The underlying principle is always that a 
conserved vector current is 1 at $q^2=0$. The problem is now
to calculate the correction to this from electromagnetism and quark masses.

For $V_{us}$ the 
vector current is $\bar s\gamma_\mu u$ so corrections
are $(m_s-m_u)^2$ not $(m_d-m_u)^2$, (Ademollo-Gatto theorem\cite{AG})
and there is a sizable
extrapolation to the zero-momentum point. For $V_{ud}$ the main problem
is that we want an extremely high precision in neutron or pion
$\beta$ decay which is hard to obtain or we have to theoretically
understand nuclear effects to a very high precision.

A $V_{us}$ determination via
hyperon $\beta$-decays has large theoretical problems. CHPT calculations
suffer from large higher orders and many parameters. Not understood
is why lowest order CHPT with model corrections works OK. References
can be found in the particle data book.
This area needs theoretical work very badly.

The experimental result
\be
\label{vus}
|V_{us}| = 0.2196\pm0.0023
\ee
together with $|V_{ud}|=0.9735\pm0.0008$ yields
\be
|V_{ud}|^2+|V_{us}|^2 = 0.9959\pm0.0026
\ee
and a mild problem for the unitarity of the CKM matrix.

The theory\cite{LR} and data behind Kaon $\beta$-decays
are both old by now. The data we expect to improve in the near future with
data from BNL, KLOE, KTeV, NA48 and possibly others.
The theory consists of oldfashioned photon loops and one-loop CHPT
calculations for the quark mass effects. The former are in the process of
being improved by Cirigliano et al. and the latter extended to
two-loops\cite{BT}. We can thus expect an improved accuracy for $V_{us}$
in the next few years.
GCHPT allows for larger quark mass effects\cite{FKS}
than the calculation
of \cite{LR}.

\section{Anomalies}
\label{anomalies}

The most celebrated result here is $\pi^0\to\gamma\gamma$.
Good agreement with the anomaly prediction
exists, at the two sigma level,
but we need to push both theory and experiment
beyond the present experimental precision of $7.8\pm0.6$~eV.

Similarly in $\eta\to\gamma\gamma$.
The discrepancy between $e^+e^-$
and  Primakoff measurements persists. KLOE should be able to contribute
significantly here.
Data for $\pi^0,\eta,\eta^\prime\longleftrightarrow \gamma^*\gamma^*$
are needed for $a_\mu$ as discussed above
and for $\eta\to\ell^+\ell^-$, even if in the latter case differences
are smaller.

$\gamma\pi\to\pi\pi$ experiment is significantly above theory,
corrections go in the right direction but only halfway
(one-loop\cite{BBC}  dispersive\cite{Holstein}).
New experiments at  JLab and possibly at COMPASS are planned.
A possible problem are the radiative corrections\cite{AKT}.
In $\eta\to\pi^+\pi^-\gamma$ the agreement between theory\cite{BBC}
and experiment is satisfactory.
Kaon decays $K\to\gamma\ell\nu,\pi\pi\ell\nu,\pi\gamma\ell\nu$
allow more tests in particular also of the sign and of the quark mass
dependence of the anomalous effects. One example is discussed below.

\section{More $\eta$}
\label{moreeta}

In $\eta$-decays there are more interesting results.
The decays $\eta\to\pi^+\pi^-\pi^0,\pi^0\pi^0\pi^0$ play a role
in the determination of the quark masses since its decay width is
$\Gamma \sim (m_u-m_d)^2 \overline\Gamma$.
The present comparison with theory is shown in Table \ref{tableeta}.
\begin{table}
\begin{center}
\begin{tabular}{|lcl|}
\hline
Exp.:      &  $280\pm28$~eV &\\
\hline
 $p^2$ & 66~eV & Ref. \cite{Cronin} \\
 $p^4$ & $167\pm50$~eV & Ref. \cite{GLeta}\\
dispersive  & $209\pm20$~eV & Ref. \cite{dispersiveeta}\\
\hline
\end{tabular}
\end{center}
\caption{Comparison of theory and experiment for $\eta\to\pi\pi\pi$.}
\label{tableeta}
\end{table}
With a slightly larger value of $m_d-m_u$, which is in fact obtained
as discussed above, we obtain good agreement.
A possible problem is that the dispersive calculations use, and can be
checked using, the Dalitz-plot parameters.
As an example, $-10^3\alpha$ in the neutral decay
was obtained in\cite{dispersiveeta} as $7\pm7$. Data
are $22\pm23$ (GAMS\cite{GAMS}),
 $52\pm18$~(Crystal Barrel\cite{CBar}) and
a preliminary result by the Crystal Ball of
$26\pm6$.
We expect more data here in the near future from KLOE, WASA and
the Crystal Ball as well as for the charged decays. Should this
discrepancy in the Dalitz plot parameters persist a reanalysis will become
necessary.

The process $\eta\to\pi^0\gamma\gamma$ provides
a good test of $p^6$ CHPT and VMD with a width prediction of
 $0.40(20)~$eV\cite{Ametller92}, present experiment gives
  $0.84(19)~$eV\cite{GAMS}.

\section{Rare semileptonic decays}
\label{semi}

Many calculations exist. See e.g. the talk by Isidori and the proceedings
of KAON99. An example is the recently improved precision on the
form-factors in $K\to\mu\nu\gamma$ by the BNL-E787 experiment\cite{E787}.
The structure dependent terms have been measured to be
\ba
|F_V+F_A|^{\mbox{exp}}&=& 0.165(0.155)\pm0.013\quad[0.14]
\nonumber\\
|F_V-F_A|^{\mbox{exp}}&=& 0.102(0.065)\pm0.085\quad[0.05]
\nonumber
\ea
The difference in the numbers with and without brackets is due to
assumptions on the $W^2$ dependence. The theory numbers in square brackets
are $p^4$ results
from \cite{BEG}. Again, improvement to the $p^6$-level is needed.

\section{Nonleptonic decays: CHPT}
\label{nonl}

This area was pioneered by \cite{KMW} and one of its major results
is the chiral symmetry tests in $K\to\pi\pi$ and $K\to\pi\pi\pi$
decays. Various relations have been produced\cite{KMW2} and
their comparison with experiment is given in
Table \ref{tablek3pi}. The overall agreement is satisfactory but the
newer CPLEAR data and effects of order $m_\pi^2$ have not been
incorporated and obviously the $\Delta I=3/2$ need experimental
improvement.
\begin{table}
$$
\begin{array}{|c|ccc|}
\hline
 & p^2 & p^4 &\mbox{experiment} \\
\hline
\alpha_1 & 74 & \mbox{input}^\dagger & 91.71\pm0.32\\
\beta_1 & -16.5  & \mbox{input}^* & -25.68\pm0.27\\
\zeta_1 & & -0.47\pm0.18^\dagger & -0.47\pm0.15\\
\xi_1 & & -1.58\pm0.19^* & -1.51\pm0.30\\
\alpha_3 & -4.1 & \mbox{input}^\bullet & -7.36\pm0.47\\
\beta_3 & -1.0 & \mbox{input}^\blacktriangle & -2.42\pm0.41\\
\gamma_3 & 1.8  & \mbox{input}^o & 2.26\pm0.23\\
\xi_3 &   & 0.092\pm0.030^\blacktriangle & -0.12\pm0.17\\
\xi_3^\prime & & -0.033\pm0.077^o & -0.21\pm0.51\\
\zeta_3 & & -0.011\pm0.006^\bullet & -0.21\pm0.08\\
\hline
\end{array}
$$
\caption{Chiral symmetry relations for various Dalitz plot
parameters in $K\to3\pi$ and their comparison with experiment.
The relations are indicated using the different superscripts.}
\label{tablek3pi}
\end{table}

In rare decays there are many successful predictions. An example is the
prediction\cite{Goity} and subsequent confirmation of $K_S\to\gamma\gamma$.
New results on this decay can be found in the KTeV and NA48 rare decay talks.

Similar predictions exist for $K_L\to\pi^0\gamma\gamma$ and
$K\to\pi\gamma^*$ and $K\to\pi Z^*$. A problem case is the
decay
$K_L\to\gamma\gamma$ which is plagued by strong cancellations but very good
predictions exists for the various $K\to\pi\nu\bar\nu$ modes.
Reviews are the talk by Isidori here and at KAON99.

\section{Nonleptonic decays and 
$\epsilon^\prime/\epsilon$}

The theoretical precision of $\epsilon^\prime/\epsilon$ is now way behind
the experimental determination reported here by KTeV and NA48.
I will only comment the approaches using chiral Lagrangian
 related methods as a main tool. In particular I concentrate
on the various analytical approaches using large $N_c$ as a guide.

The major problem is that due to the virtual $W$-bosons
we need an integration over {\em all} scales.
The short-distance part can be resummed using renormalization group
methods and is known to two-loops, calculated by
two groups who are in perfect agreement, see
\cite{Buras} for recent lectures and references.

The large $N_c$ approach combined with CHPT
was pioneered by Bardeen et al.\cite{BBG}
and is now used by several groups.
The Dortmund\cite{Dortmund} group uses pure CHPT at long
distances and matches it directly onto short-distance QCD,
the Granada-Lund\cite{BP} group uses the $X$-boson method to match
correctly onto the short-distance schemes and the ENJL model
to improve on the hadronic picture at intermediate scales.
Finally the Valencia group\cite{Valencia} uses a 
semi-phenomenological
approach to calculate $\epsilon^\prime/\epsilon$.
It should be noted that the Dortmund and Granada-Lund groups also
reproduce within the uncertainties of their approach the $\Delta I=1/2$
rule which no other method can claim so far.

Besides these calculations which fully predict $\epsilon^\prime/\epsilon$
there has been major progress on several fronts recently.
In particular it has been realized that $B_7$ and $B_8$ can be studied
in the chiral limit from $\tau$-decay data.\cite{B7B8,B7B82}
All groups basically agree in the method but differ
in the treatment of scheme-dependence and error estimates due to the
spectral functions from  $\tau$ data.
Comparison of these results with the calculations mentioned above is
in.\cite{B7B82}
Similar dispersive results for the $B_K$
parameter\cite{PRBK}
were in reasonable
agreement with the equivalent Granada-Lund one\cite{BP,BKBP}.

Calculations with no good theory for the long-distance--short-distance
matching have not been included here. In particular I included neither
the factorization (Taipei)
nor the chiral quark model (Trieste\cite{Trieste}) results.

The estimates of $\epsilon^\prime/\epsilon$
shown in Table~\ref{tableepsp}, increased significantly
due to three effects:
\begin{itemize}
\item $\Omega_{IB }$ has gone down, removing part
of the cancellations due to isospin breaking.\cite{EPN}
\item Final state interactions should be included
both in the real and in the imaginary parts.\cite{Trieste}
Typically the real part
was treated experimentally and the imaginary part to order $p^2$.
Correcting this produces a strong enhancement.\cite{Trieste,PP}
Criticisms now focus on the {\em size}
 of this effect not its {\em existence}.
\item Large, not fully understood,
non-factorizable corrections\cite{BP} to $B_6$.
\end{itemize}
\begin{table}
\begin{center}
\begin{tabular}{|lc|}
\hline
  & { $10^4 \epsilon^\prime/\epsilon$}\\
\hline
Granada-Lund   & $34\pm18$\\
Dortmund       & $11\pm 5$\\
Valencia       & $17\pm9$ \\
Data           & $17.2\pm1.8$\\
\hline
\end{tabular}
\end{center}
\caption{Some predictions for $\epsilon^\prime/\epsilon$ and the 
the data.}
\label{tableepsp}
\end{table}
\section{Baryons}

Baryons, hyperons and nuclei are also areas where chiral Lagrangians
play an important role. The one baryon sector was reviewed
in,\cite{onebaryon} many more results can be found in
 the proceedings of Chiral Dynamics 2000
at JLab.

\begin{figure}
\includegraphics[width=\columnwidth]{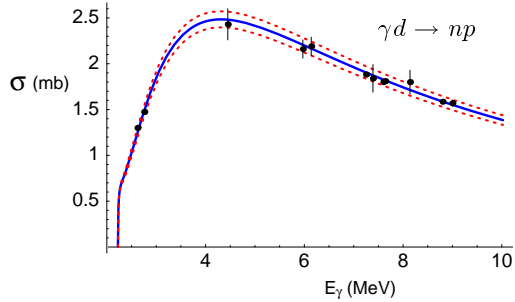}
\caption{The cross-section relevant for deuteron breakup
from EFT, figure from \rcite{Chen}.}
\label{fig1gd}
\end{figure}

In the two or more nucleon sector some recent progress has been obtained
by the advent of a proper power-counting\cite{KSW}. Many applications
can be found in the abovenamed proceedings and in the
review.\cite{twobaryon} I only quote two examples relevant for cosmology
and solar physics. The EFT calculation\cite{Chen}
allowed to reduce the uncertainty for deuteron break-up in the early
universe from 5\% to 1\%. More importantly, good data at one point
will improve the prediction over the whole relevant energy range.
The cross-section is shown in Fig. \ref{fig1gd}.
For the cross-section of the main solar proton fusion process $pp\to d e^+\nu$
EFT methods allow a one parameter prediction.\cite{ppd}
A 10\% measurement of $\bar\nu_e d$ scattering at a
reactor allows a 3\% prediction of the $pp$ rate.

\section{Colour Superconductivity}

At very high baryon densities a new phase of QCD is expected to
appear\cite{CFL1} where we get diquark condensation in 
the anti triplet colour channel of $3_c\times 3_c$
rather than in the colour singlet quark-antiquark in $3_c\times \bar 3_c$
of the usual QCD vacuum. This phase has colour-flavour locking due to the
nature of the condensate
and the symmetries are broken in the pattern
\ba
SU(3)_c\times SU(3)_L \times SU(3)_R \times U_V(1)
\nonumber\\
\Rightarrow SU(3)_{c+L+R}\times Z_2(q\to-q)
\ea
This results in
8 massive gluons, 8 Left-Right Goldstone Bosons, one Baryon-number
Goldstone Boson. This looks very much like
the spectrum of  8 vectors and 8 pseudoscalars in the usual QCD vacuum.
Indeed the whole formalism of EFT can also be used here as can be seen in
the review by Alford.\cite{Alford}
Some peculiarities are that the
mass spectrum is quadratic in the quark masses due to the
 $Z_2$ symmetry of the condensate and that instead of the more familiar
photon-rho mixing we now have photon gluon mixing.\cite{photongluon}

\section{Not covered}

A few more areas with recent progress are
\begin{itemize}
\item
The question of large $N_f$ and GCHPT.\cite{ST}
\item
$\eta$-$\eta^\prime$ mixing.\cite{Kaiser}
\item
Phenomenological Chiral Lagrangians in resonance decays especially
$\phi$-decays and $\tau$ phenomenology.\cite{contrib}
\item 
CHPT in the single baryon sector.\cite{onebaryon,becher}
\item
CHPT for vector mesons.\cite{vector}
\item
Resummation work beyond the ones discussed here.\cite{Oset}
\item
Connection with lattice QCD: the use of CHPT to extrapolate lattice
results to physical
quark masses and study effects of (partial) quenching and finite volume.
\end{itemize}

\section{Conclusions}

I hope to have convinced you that the
field of Chiral Lagrangians is very active and has many applications
with relevance  for a wide spectrum of phenomena in high-energy
and nuclear physics.
\section*{Acknowledgments}
This work is supported by the Swedish Research Council and
EU-TMR network, EURODAPHNE,
Contract No. ERBFMRX--CT980169.

\end{document}